\documentclass{article}

\usepackage{arxiv}
\usepackage[utf8]{inputenc} 
\usepackage[T1]{fontenc}    
\usepackage{hyperref}       
\usepackage{url}            
\usepackage{booktabs}       
\usepackage{lipsum}		
\usepackage{graphicx}
\usepackage{natbib}
\usepackage{doi}
\usepackage{subfig}
\title{Predicting Swarm Equatorial Plasma Bubbles via Machine Learning and Shapley Values}

\date{} 					
\author{ \href{https://orcid.org/0000-0003-1237-6518}{\includegraphics[scale=0.06]{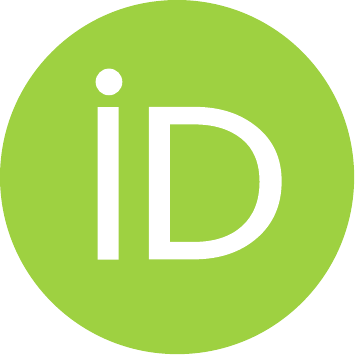}\hspace{1mm}Sachin A. ~Reddy}\\
	Mullard Space Science Laboratory\\
	University College London\\
	Dorking, UK \\
	\texttt{sachin.reddy.18@ucl.ac.uk} \\
	\And
	\hspace{1mm}Colin Forsyth \\
	Mullard Space Science Laboratory\\
	University College London\\
	Dorking, UK \\
	\texttt{colin.forsyth@ucl.ac.uk} \\
	\And
	\hspace{1mm}Anasuya Aruliah \\
	Dept. of Physics and Astronomy\\
	University College London\\
	London, UK \\
	\texttt{a.aruliah@ucl.ac.uk} \\
	\And
	\hspace{1mm}Andy Smith \\
	Mullard Space Science Laboratory\\
	University College London\\
	Dorking, UK \\
	\texttt{andy.w.smith@ucl.ac.uk} \\
	\And
	\hspace{1mm}Jacob Bortnik \\
	Dept. of Atmospheric and Oceanic Studies \\
	University of California at Los Angeles (UCLA)\\
	Los Angeles, CA, USA \\
	\texttt{jbortnik@atmos.ucla.edu} \\
        \And
	\hspace{1mm}Ercha Aa \\
	Haystack Observatory \\
    Massachussets Institute of Technology (MIT) \\
	Cambridge, MA, USA \\
	\texttt{aercha@mit.edu} \\
  	\And
	\hspace{1mm}Dhiren O. Kataria \\
	Southwest Research Institute\\
	San Antonio, TX, USA\\
	\texttt{dhirendra.kataria@swri.org} \\
	\And
	\hspace{1mm}Gethyn Lewis \\
	Mullard Space Science Laboratory\\
	University College London\\
	Dorking, UK \\
	\texttt{g.lewis@ucl.ac.uk} \\
}

\begin{document}
\maketitle

\begin{abstract}
In this study we present AI Prediction of Equatorial Plasma Bubbles (APE), a machine learning model that can accurately predict the Ionospheric Bubble Index (IBI) on the Swarm spacecraft. IBI is a correlation ($R^2$) between perturbations in plasma density and the magnetic field, whose source can be Equatorial Plasma Bubbles (EPBs).
EPBs have been studied for a number of years, but their day-to-day variability has made predicting them a considerable challenge.
We build an ensemble machine learning model to predict IBI. We use data from 2014-22 at a resolution of 1sec, and transform it from a time-series into a 6-dimensional space with a corresponding EPB $R^2$ (0-1) acting as the label.
APE performs well across all metrics, exhibiting a skill, association and root mean squared error score of 0.96, 0.98 and 0.08 respectively. The model performs best post-sunset, in the American/Atlantic sector, around the equinoxes, and when solar activity is high. This is promising because EPBs are most likely to occur during these periods.
Shapley values reveal that F10.7 is the most important feature in driving the predictions, whereas latitude is the least. The analysis also examines the relationship between the features, which reveals new insights into EPB climatology.
Finally, the selection of the features means that APE could be expanded to forecasting EPBs following additional investigations into their onset.

\textbf{This work has now been published \citep{reddy2023predicting}, please use the following reference if you cite the paper:}

Reddy, S. A., C. Forsyth, A. Aruliah, A. Smith, J. Bortnik, E. Aa, D. O. Kataria, and G. Lewis. "Predicting swarm equatorial plasma bubbles via machine learning and Shapley values." Journal of Geophysical Research: Space Physics (2023): e2022JA031183.
\end{abstract}

\keywords{Equatorial Plasma Bubbles \and Supervised Machine Learning \and Shapley Values}

\section{Introduction}

In the post sunset \textit{F} region of the ionosphere, plumes of low density plasma, known as \textit{Equatorial Plasma Bubbles} (EPBs) are prone to form. These bubbles were first observed in ionosonde traces, and have subsequently been captured by radar, air glow images, and in-situ detectors \citep{Woodman1976, Argo1986, Retterer2014}. EPBs can cause fluctuations in the amplitude and phase of radio waves that traverse through them \citep{Kintner2007}. These \textit{scintillations} adversely affect Global Navigation Satellite System (GNSS) and other communication systems which rely on quiet ionospheric conditions. Their morphology, onset, and development is complex and has been the subject of numerous studies over the years.

In the sunlit hemisphere, the neutral wind generally travels in an easterly direction towards the day-night terminator \citep{Heelis2012}, forcing the plasma in an upwards zenith direction under the action of the Lorentz force. Once in the nightside, ionization ceases and recombination dominates. This leads to a large density gradient between the \textit{E} and \textit{F} regions. When the interface between these layers is perturbed, the rarefied \textit{E} / lower \textit{F} layers are forced vertically upward into the higher density plasma, which itself is being pulled down under the action gravity \citep{Kelley2009}. This mechanism is known as a \textit{Generalized Rayleigh-Taylor instability}, $\gamma$, and its growth rate is described by \citep{Sultan1996}
The growth rate of the RTI, $\gamma$,  was formulated by Sultan in 1996

\begin{equation}
    \gamma=\frac{\sum_{P}^{F}}{\sum_{P}^{E}+\sum_{P}^{F}}\left(V_{P}-U_{L}^{P}-\frac{g_{e}}{v_{ef f}}\right) K^{F}-R,
    \label{eq:sultan-rt}
\end{equation}

where $\sum_{P}$ is the flux integrated Pederson conductivity for the E and F layers, $V_p$ is the vertical plasma drift, $U_{L}^{P}$ is the Pederson conductivity weighted neutral meriodional wind, ${g_e}$ is the altitude corrected acceleration due to gravity, $v_{eff}$ is the ion-neutral collision frequency, $K^F$ is the total \textit{F} region flux electron tube content, and $R$ is the ion-electron recombination rate \citep{Sultan1996}. Because of the conductivity ratio of $\sum^{F}_P / \sum^{E}_P + \sum^{F}_P$ the onset of an EPB can only occur at night when the \textit{E} region conductivity is very weak. High values for $\sum_P^E$, $U_{L}^{P}$, and $R$ will act to suppress an EPB, whereas high values of $V_p$, $K^F$ and, $g_e/v_{eff}$ will destabilize the plasma and enhance the likelihood of an EPB \citep{Sultan1996, Carter2020}.

The spatiotemporal prediction of EPB occurrence has remained an on-going challenge for a number of years. Whilst the growth rate is described by equation (\ref{eq:sultan-rt}), the terms themselves are influenced by local time, geolocation, season, and solar and geomagnetic activity \citep{Burke2004, Carter2014, Kumar2016, Smith2017, Aa2020, Carter2020}. To complicate matters, these climatological markers can often contradict themselves and the relationship between them is nuanced. 
Geomagnetic activity can both enhance and suppress the onset of an EPB via modified equatorial electrodynamics due to different perturbation electric fields \citep{Aa2019, Abdu2012, Carter2016, Kumar2016}. The under-shielding prompt penetration electric field (PPEF) tends to be dominant during the storm main phase due to suddenly varying magnetosphere convection, which has an eastward polarity in the dayside through local dusk but westward polarity in the nighttime. This typically enhances equatorial upward plasma drift in the dusk sector and thus facilitates the development of postsunset EPBs, but may disrupt post-midnight EPBs via downward plasma drift. On the other hand, the disturbance dynamo electric field (DDEF) -- due to changes in global thermosphere circulation -- usually dominates during the storm recovery phase, which has an opposite polarity with PPEF and so tends to suppress postsunset EPBs, but enhances postmidnight EPBs. In addition, the over-shielding penetration electric field due to substorm activity has an opposite polarity with that of PPEF, thereby suppressing the postsunet EPBs, but enhancing postmidnight EPBs. The combination and interaction of these perturbation electric fields leads to complicated occurrence patterns and spatio-temporal variations of EPBs.

Interest in machine learning (ML) within the heliophysics community has grown enormously in recent years \citep{Camporeale2019}, but its direct application to EPBs remains more limited. A \textit{random forest regressor} has been employed to predict the vertical plasma drifts, or $V_P$ in equation (\ref{eq:sultan-rt}) \citep{Shidler2020}. This is a significant term in the overall onset of an EPB \citep{Tsunoda2018}. Others have used an all-sky imager to train a \textit{convolution neural network} to detect EPBs, although the results seem more preliminary \citep{Srisamoodkham2022}. EPBs are also known as \textit{Spread F}, which is a broader class of irregularities or wave-like structures within the ionosphere \citep{Lan2018}. Here ensemble and \textit{deep learning} methods have been employed to classify and automatically detect Spread F in ionograms \citep{Lan2018, Luwanga2022}. EPBs are a known cause of radio wave scintillations \citep{Kintner2007}, and ML has been used to predict when and where scintillations may occur \citep{Jiao2017, Linty2018, Mcgranaghan2018}. Lastly, deep learning has also been applied to predict storm-driven irregularities within the ionosphere \citep{Liu2021}.

In this study we present AI Prediction of EPBs (APE), an ML model that predicts the Ionospheric Bubble Index (IBI) index on Swarm. First we introduce Swarm and the IBI product. Then, we analyze the $R^2$ value which is created by IBI and contains plasma bubbles. Thirdly, we describe the ML models and their performance. Finally, we use Shapley values to interpret and explain the complex interactions within APE, all of which highlights the scientific benefits of using such an approach.

\section{Instrumentation, Data and Observations}

Swarm is a three-spacecraft Earth exploration constellation that launched on 22 November 2013. Two spacecraft, \textit{Alpha} and \textit{Charlie}, were at an initial altitude of roughly 470 km, whereas \textit{Bravo} was at 520 km \citep{Friis2008}. Alpha and Charlie operate side-by-side, separated by about 1.4° in longitude. All three have a circular near-polar orbit of 87°. Swarm automatically detects EPB’s via its Ionospheric Bubble Index (IBI) product, which we use to train our machine learning models. EPBs can be characterised by prolonged and simultaneous changes in B and Ne \citep{Stolle2006}. Swarm has an on-board magnetometer and Langmuir probe to measure these quantities respectively. IBI correlates the strength of $\Delta$Ne and $\Delta$B (where residual B fluctuations in the range 0.04 -0.5Hz exceed 0.2 nT) using the Pearson correlation co-efficient (R). An $R^2$ $>$ 0.5 is tagged as a ‘confirmed bubble’ and $<$ 0.5 is an ‘unconfirmed bubble’. In addition to a strong $R^2$ score, bubbles are only confirmed if: detected at night, at latitude $<$ 45°, there are no gaps in the data, and no non-physical measurements from the Langmuir probe or magnetometer. This reduces the risk of contamination from non-EPB events, but it does not stop some plasma blobs from being erroneously labelled as EPBs \citep{Park2013}. These will be more pronounced during solar minimum \citep{Choi2012}.

\begin{figure}[htp]
    \centering
    \includegraphics[width=\textwidth]{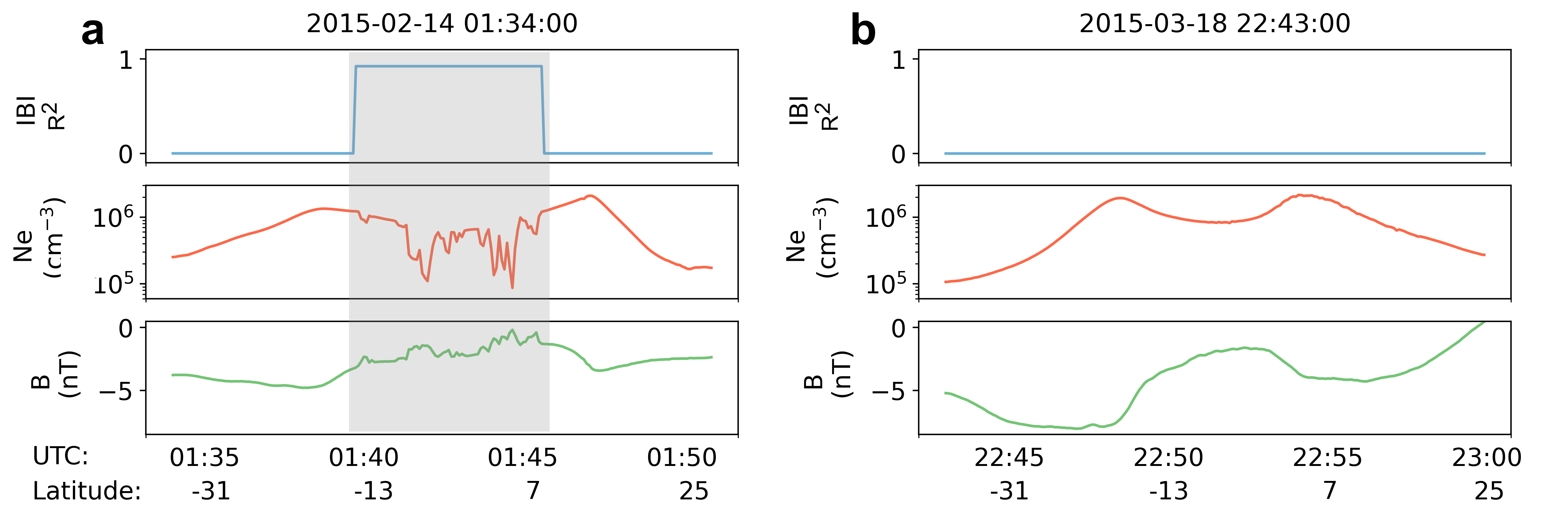}
    \caption{Two examples of Swarm passing over the equator. \textbf{a}) Swarm detects an EPB as indicated by the grey-box.
    \textbf{b}) Quiet time conditions with no bubbles present.}
    \label{fig:ins_epb-no-epb}
\end{figure}

An example IBI EPB is shown inside the grey box of Figure \ref{fig:ins_epb-no-epb}a. Here a $\Delta$B occurs simultaneously with a $\Delta$Ne between the period 0140 – 0147, which in turn triggers an IBI $R^2$ of 0.97. This value equates to a very high chance of EPB detection. A quiet bubble-free ionosphere is shown in Figure \ref{fig:ins_epb-no-epb}b. 

\begin{figure}[htp]
    \centering
    \includegraphics[width=11cm]{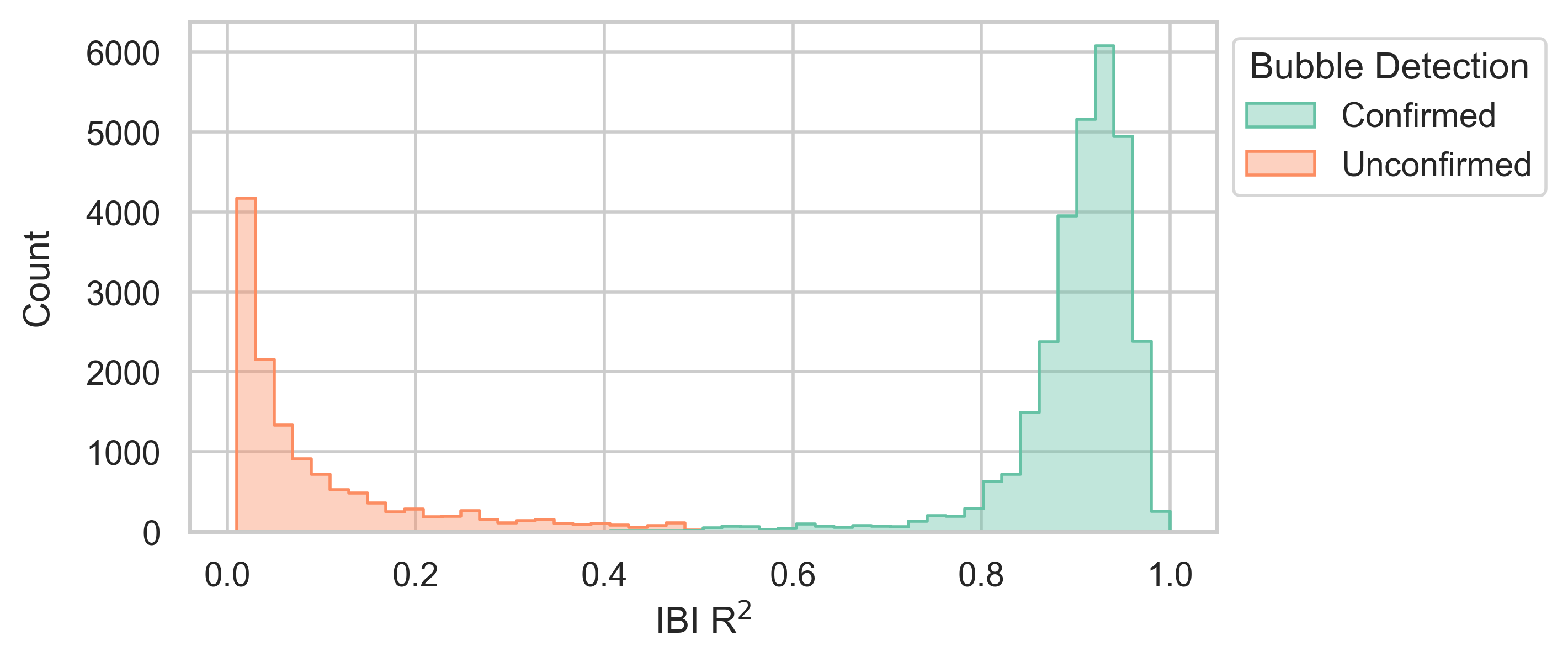}
    \caption{Distribution of $R^2$ detected by IBI across 2014-2022, where $R^2 >$ 0.5 = `confirmed', and $R^2$ $<$ 0.5 = `unconfirmed' \citep{Park2013}.}.
    \label{fig:ins_prob-dis}
\end{figure}

IBI data was accessed via ESA's virtual research environment for Swarm \url{(https://vires.services)} and the Python package \textit{viresclient} \citep{Smith2022}. We also use viresclient to map F10.7 and Kp values to the IBI dataset. We use data from 2014 - 2022 at a resolution of 1s across all three spacecraft where $R^2 >$ 0. The date range covers the declining phase of solar cycle 24 and the start of solar cycle 25. We transform the data from a time-series into a 6-dimensional space consisting of MLT, latitude, longitude, day-of-the-year, Kp, and F10.7, with each dimension having a corresponding $R^2$ value (0-1) provided by IBI. This allows us to make a prediction of IBI based on the climatology of EPBs which are dependent on time, geolocation, season, and geomagnetic and solar activity \citep{Burke2004, Carter2016, Carter2020, Aa2020}. It also ensures that the model can be expanded to \textit{forecasting}, as Kp and F10.7 are readily available via NOAA \url{(https://www.swpc.noaa.gov/products)}.
After re-binning and cleaning, we have $\sim$42k samples for the machine learning models.  Figure \ref{fig:ins_prob-dis} shows the distribution of $R^2$ across the 9-year period. As seen the majority of values cluster around $R^2 = 0$ and $R^2=0.9$. We are mainly interested in $R^2>0.7$. 

\begin{figure}[htp]
    \centering
    \includegraphics[width=\linewidth]{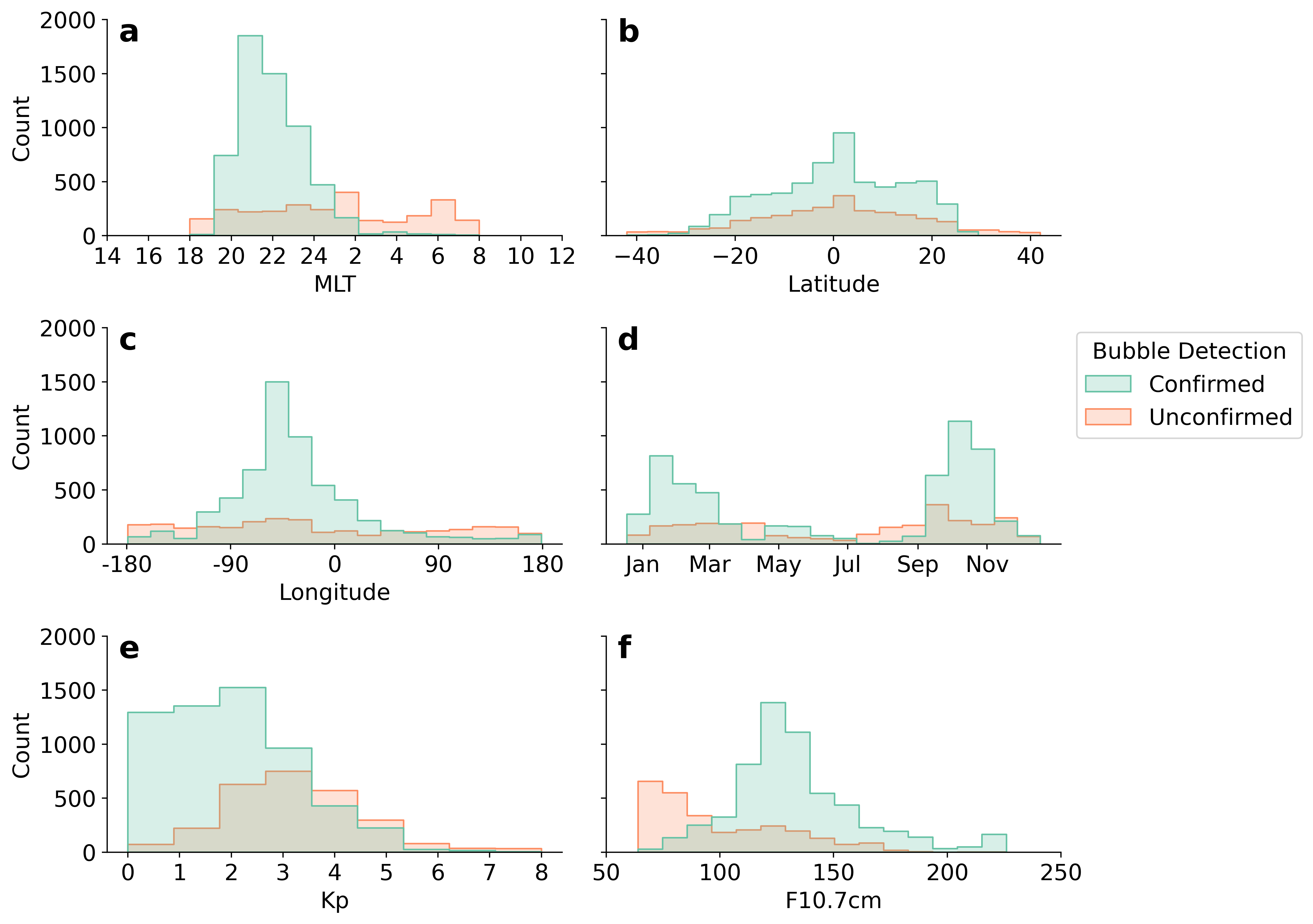}
    \caption{Bubble detection across the six climate features. \textit{Confirmed} bubbles ($R^2 >$ 0.5) exhibit different spatio-temporal characteristics than \textit{unconfirmed} magnetic-only fluctuations  ($R^2 <$ 0.5).}
    \label{fig:ins_clim-hists}
\end{figure}

Next, we examine the distribution of the 42k samples across the 6 features. Figure \ref{fig:ins_clim-hists} shows that `confirmed' and `unconfirmed' bubbles are not uniform across the climate markers.

Most confirmed bubbles are in the post-sunset time frame (19-24 MLT), with a small increase at 4 MLT (Fig. \ref{fig:ins_clim-hists}a).
The distribution of confirmed bubbles is centered around the geographic equator with only a few instances beyond 25° glat (Fig. \ref{fig:ins_clim-hists}b).
Next, we see that most bubbles occur in the American/Atlantic sector, but that instances exist at all longitudes (Fig. \ref{fig:ins_clim-hists}c).
The majority of EPBs occur around the equinox months and winter solstice, with little activity in July and August (Fig. \ref{fig:ins_clim-hists}d).
Fig. \ref{fig:ins_clim-hists}e shows that the number of confirmed EPBs declines with Kp, and there are no bubbles detected at Kp $>$ 7. 
Lastly, we see that EPB activity peaks around F10.7 = 125, but an additional population exists at F10.7 = 220 (Fig. \ref{fig:ins_clim-hists}f). This panel also reveals that EPBs are generally less likely to occur at F10.7 $<$ 90.
Overall, these results align with the existing literature on EPB climatology \citep<e.g.,>[]{Burke2004, Abdu2012, Park2013, Carter2016, Aa2020}. Figure \ref{fig:ins_clim-hists} also provides some insight into magnetic-only fluctuations (R$^2 < $ 0.5)  in the ionosphere, with F10.7 and Kp showing some interesting distributions (Fig. \ref{fig:ins_clim-hists}e-f).

\section{Machine Learning}
We use supervised machine learning (ML) algorithms to predict the IBI value provided by Swarm. Supervised methods require \textit{labels}, $y_i$, which we assign to $R^2$.  We use regression specific architectures as the labels are considered a continuous value. ML has a unique ability to identify complex relationships in data that contains rare events. It can also handle heterogeneity in space-time and large amounts of noise \citep{Karpatne2018, Camporeale2019}. Because of this, we believe it is well suited to the task of predicting IBI and the EPBs contained within it. 

Our main algorithm is the \textit{eXtreme Gradient Boosting} (XGBoost) method which is a tree-based ensemble learner. XGBoost has good control over bias and variance, whilst remaining computationally inexpensive to train and enabling \textit{explainability} \citep{Chen2016, Lundberg2020}. The model's prediction ability is expressed by

\begin{equation}
    \hat{y_i}=\sum_{k=1}^{K}f_k(x_i), f_k \in \mathcal{F},
    \label{eq:xgb-model}
\end{equation}

where $\hat{y_i}$ is the prediction value, $K$ is the number of trees, $x_i$ is the input data, $f_k$ is a function in the functional space $\mathcal{F}$, and $\mathcal{F}$ is the set of all the possible regression trees \citep{Chen2016}. 
To evaluate the model's performance we need an \textit{objective function} \citep{Geron2019}

\begin{equation}
    obj =\sum_{i=1}^n l(y_i, \hat{y_i}^{(t)}) + \sum_{k=1}^K \omega (f_k),
    \label{eq:xgb-obj-func}
\end{equation}

where $y_i$ is the target value ($R^2$), $\hat{y_i}^{(t)}$ is the prediction of the $i^{th}$ instance at the $t^{th}$ iteration, and $\omega$ is the complexity of the model \citep{Chen2016}. The term on the left is the \textit{loss function}, and the term on the right is the \textit{regularization} term. Regularization controls the magnitude of the parameters, and thus reduces the model's complexity \citep{Geron2019}. We use the XGBoost package for python \url{(xgboost.readthedocs.io)} and Sci-kit learn \url{(scikit-learn.org)}  to perform the modelling and analysis. \textit{GridSearchCV} was used to identify the optimal hyperparameters, which are as follows: \textit{estimators = 300}, \textit{alpha = 0.1},  \textit{subsample = 0.5}, and \textit{eta = 0.2}. The last three parameters are used to prevent overfitting. We divide the samples into train and test datasets with a 80\%-20\% split. This is randomised initially and then fixed to prevent data leakage across the training runs. 

We also tested a Random Forest method \citep{Breiman2001} and a standard linear regression approach as part of our study. These will feature as a basis for global performance comparison, but are not subject to extensive analysis.

\begin{figure}[htp]
    \centering
    \includegraphics[width=10cm]{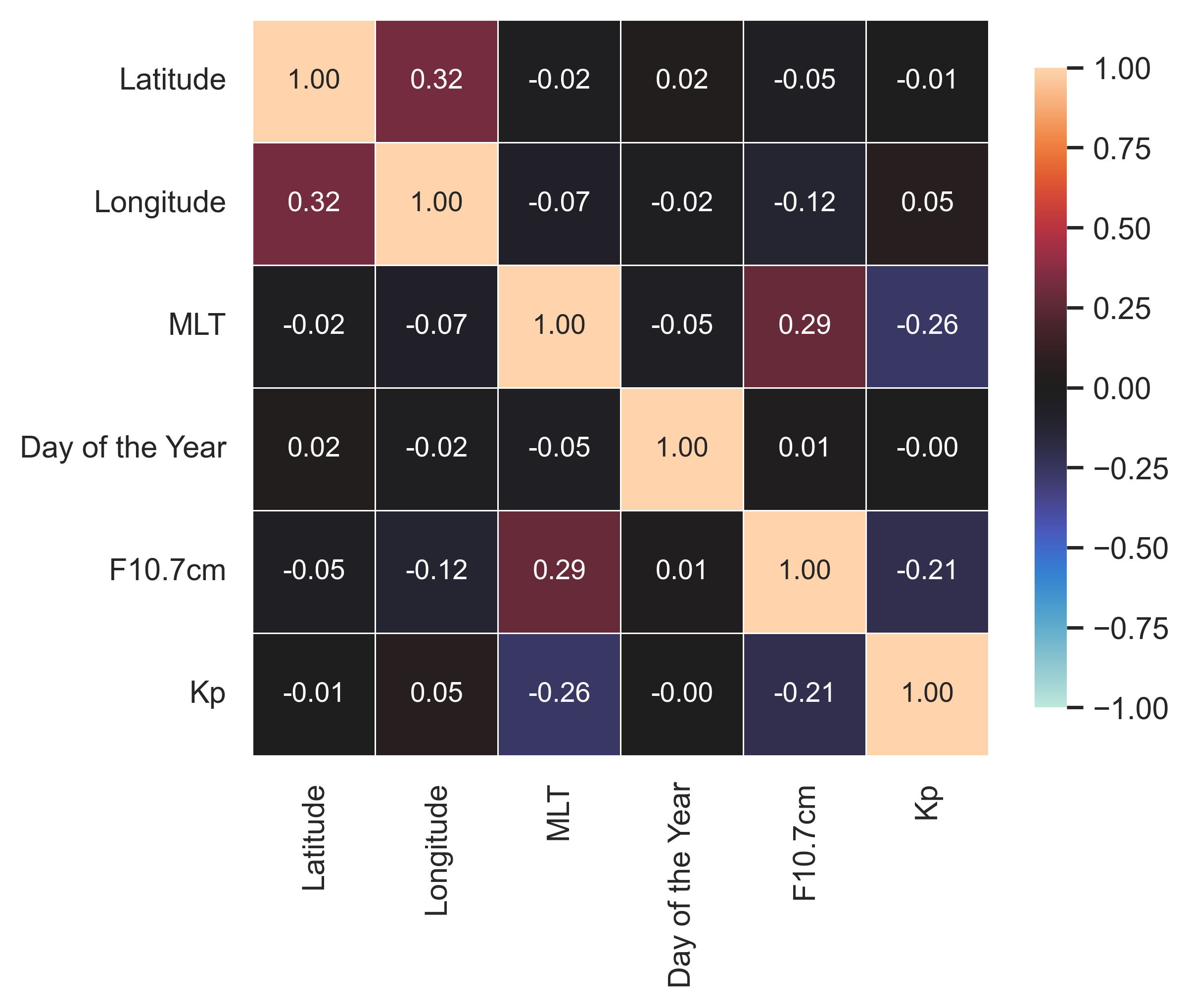}
    \caption{A correlation plot showing the relationship between the features. No strong linear correlation exists between any of the features.}
    \label{fig:ml_corr-plot}
\end{figure}

The model's input features and the linear correlation between them is shown in Figure \ref{fig:ml_corr-plot}. It reveals that there is no strong \textit{linear} correlation between any of the features, which provides further justification for using an ML approach.

\subsection{Assessment Metrics}
Several metrics are used to assess the performance, skill, and association of the model. \textit{Root Mean Squared Error} (RMSE) and \textit{Mean Absolute Error} (MAE) are typical performance tests for regression problems \citep{Chai2014},

\begin{equation}
    \mathrm{MAE}= \frac{1}{n} \sum_{i=1}^{n}\left|(\hat{y_i} - y_i )\right|,
    \label{eq:mae}
\end{equation}

\begin{equation}
    \mathrm{RMSE}=\sqrt{\frac{1}{n} \sum_{i=1}^{n} (\hat{y_i} - y_i )^{2}},
    \label{eq:rmse}
\end{equation}

where $n$ is the number of samples. Accuracy metrics tell us how close the prediction is to the true value, but they do not tell us how well the model captures the up-and-down trends of the dataset. \textit{Association} can be represented by the Pearson correlation coefficient $R$

\begin{equation}
    R=\frac{\sum\left(y_i-\bar{y}\right)\left(\hat{y_i}-\bar{\hat{y}}\right)}{\sqrt{\sum\left(y_i-\bar{y}\right)^2 \sum\left(\hat{y_i}-\bar{\hat{y}}\right)^2}},
    \label{eq:R-assoc}
\end{equation}

This tells us if the predictions are close to the target in some part of the data range, but not in others. An ideal value is $R=1$. Finally, we examine the skill of the model by looking at its \textit{Prediction Efficiency} which is based on its mean square error \citep{Murphy1988}

\begin{equation}
    P E=1-\frac{\sum\left(\hat{y_i}-y_i\right)^2}{\sum\left(y_i-\bar{y}\right)^2},
    \label{eq:pe-skill}
\end{equation}

A model with perfect skill is $PE = 1$, while $PE < 0$ shows that the model is no better at making predictions that the average of the target values $\langle y \rangle$. 

\section{Results}
The following section presents the performance of the machine learning models in terms of error, association, and skill. It goes on to interpret the behavior of the XGBoost model via Shapley values, determining the importance of the features and the relationships between them. 

Figure 5a shows the association (Eq. \ref{eq:R-assoc}) and skill (Eq. \ref{eq:pe-skill}) of the three modelling techniques. As shown, the machine learning techniques outperform the standard linear model, particularly with respect to prediction efficiency (0.45 vs. 0.96), which justifies their use. The same trend continues with RMSE (Eq. \ref{eq:rmse}) and MAE (Eq. \ref{eq:mae}), with the RF and XGBoost architecture outperforming the linear regression method across both metrics. The ensemble learners offer a considerable leap across the four metrics, but XGBoost comfortably outperforms the RF in all areas. It achieves a PE, R, MAE, and RMSE of 0.96, 0.98, 0.05, and 0.08 respectively, all of which are excellent scores. XGBoost also trains 3.8x faster than the RF, because it sub-samples and approximates the split points amongst the trees\citep{Chen2016}. We now select the XGBoost model for further analysis and name it \textit{AI Prediction of EPBs}, or APE.

\begin{figure}
    \includegraphics[width=\linewidth]{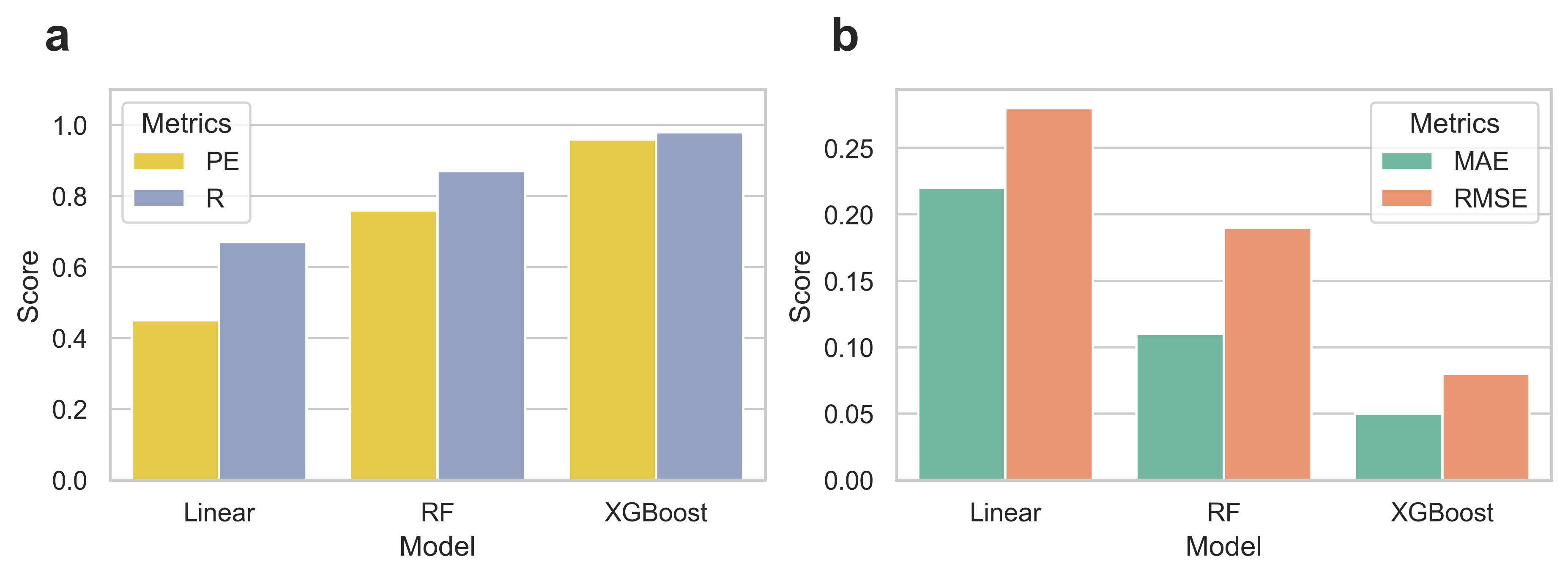}
    \caption{The skill (PE), association (R), and performance (MAE, RMSE) of three learning models on the 20\% test set. XGBoost outperforms the random forest and linear method across all four metrics.}
    \label{fig:results_skill-performance}
\end{figure}

\subsection{APE}
\begin{figure}[htp]
    \centering
    \includegraphics[width=\linewidth]{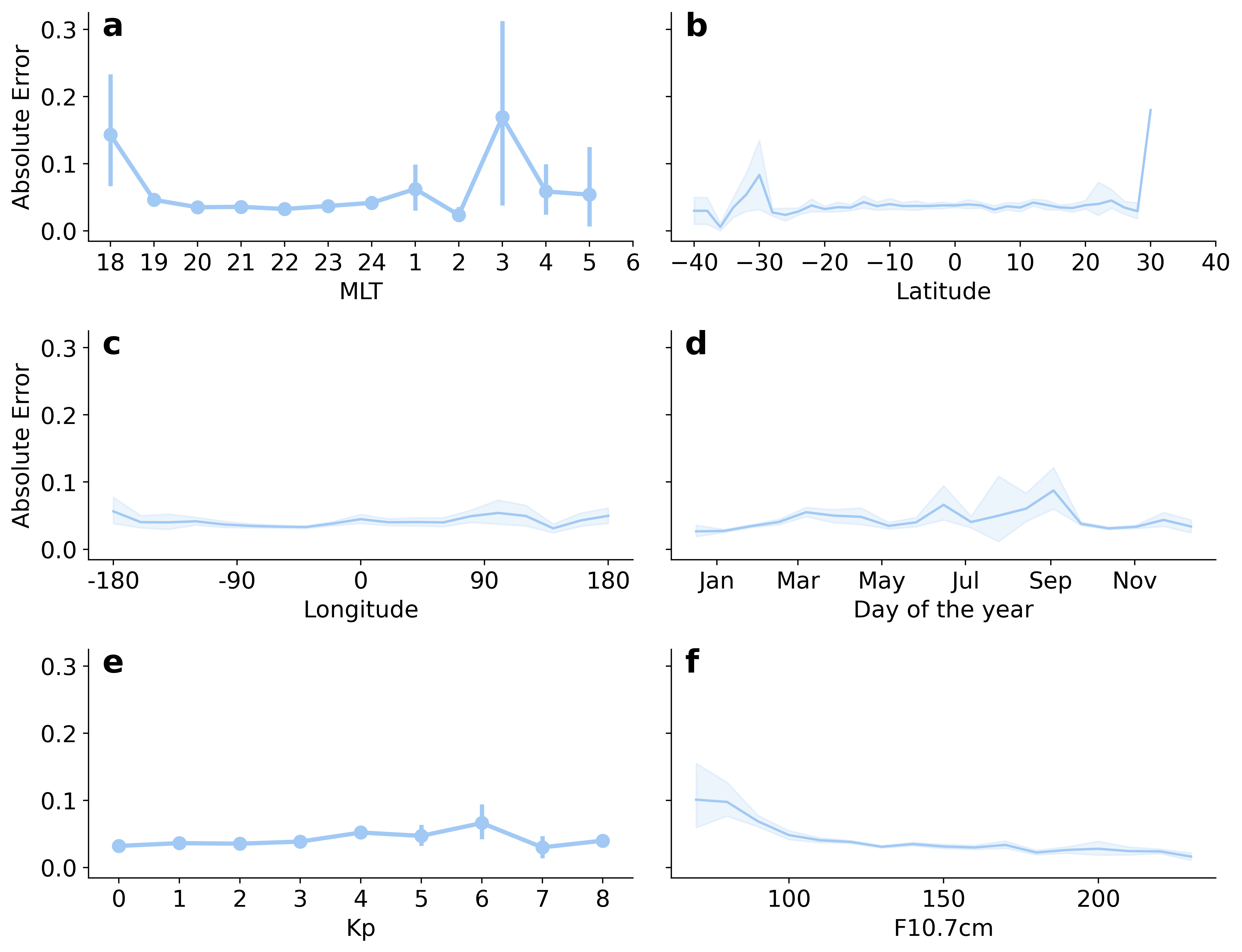}
    \caption{The \textit{absolute error} $|(\hat{y_i} - y_i)|$ of APE across the 6 climate features. 0 is an ideal score. The uncertainties are calculated with the \textit{bootstrapping} method \citep{Efron1994}, and are represented by vertical bars (a\&e) and the shaded areas (b-d, f).}
    \label{fig:results_abs-err}
\end{figure}

Figure \ref{fig:results_skill-performance} tells us how APE is performing at a global level, but it does not tell us how it performs across the feature space. For example, does the model perform better at certain local times or during specific levels of geomagnetic activity? Figure \ref{fig:results_abs-err} looks at the \textit{absolute error} between the prediction and tagret, $|(\hat{y_i} - y_i)|$, across the features. Error bars are calculated using the \textit{bootstrapping} method \citep{Efron1994}.

Generally speaking, APE performs very well across the entire feature space (Fig. \ref{fig:results_abs-err}). It performs poorer at 18 and 3 MLT (Fig. \ref{fig:results_abs-err}a), outside the equatorial region (Fig. \ref{fig:results_abs-err}b), and during low F10.7 (Fig. \ref{fig:results_abs-err}f). These are periods when EPB activity is expected to be lower and is therefore not of concern. The performance also tracks directly to the availability of the data (Fig. \ref{fig:ins_clim-hists}). That is, when there are more confirmed EPB events to learn from, model performance increases.

\subsection{Explainability}
A key tenet of the study is to understand the factors that influence predictions, as well as the connections between them. To do this we use \textit{Shapley Values}, which allow us to approximate feature contribution via cooperate game theory \citep{shapley1953stochastic}. The \textit{SHapley Additive exPlanations} (SHAP) package for Python \url{(shap-lrjball.readthedocs.io/)} treats the features as players, and prediction of R$^2$ as the pay-off \citep{Lundberg2020}. The predictions and SHAP contributions are calculated with

\begin{equation}
    f(x) = E[f(x)] + \sum_n{\phi_n},
    \label{eq:shap}
\end{equation}

where $f(x)$ is the prediction of $R^2$, $E[f(x)]$ is the \textit{expected value} which is $\approx \langle R^2 \rangle$ and is equal to 0.66, and $\phi_n$ is the SHAP value for each of the features $n$. $\phi$ represents the contribution to the pay-off, weighted and summed over all possible feature value combinations. Shapley values have the properties of efficiency, symmetry, and additivity, which ensures the pay-off is \textit{fair} \citep{shapley1953stochastic, Lundberg2020}. $E[f(x)]$ can be thought of as the climatology of $R^2$, and each of the feature values can contribute to this in a positive ($\phi > 0$) or negative ($\phi < 0$) way. Shapley values are emerging as the de facto method for explaining the output of ML models \citep{Merrick2020}, but their interpretation requires caution and expertise \citep{Kumar2020}.

\begin{figure}[htp]
    \centering
    \includegraphics[width=10cm]{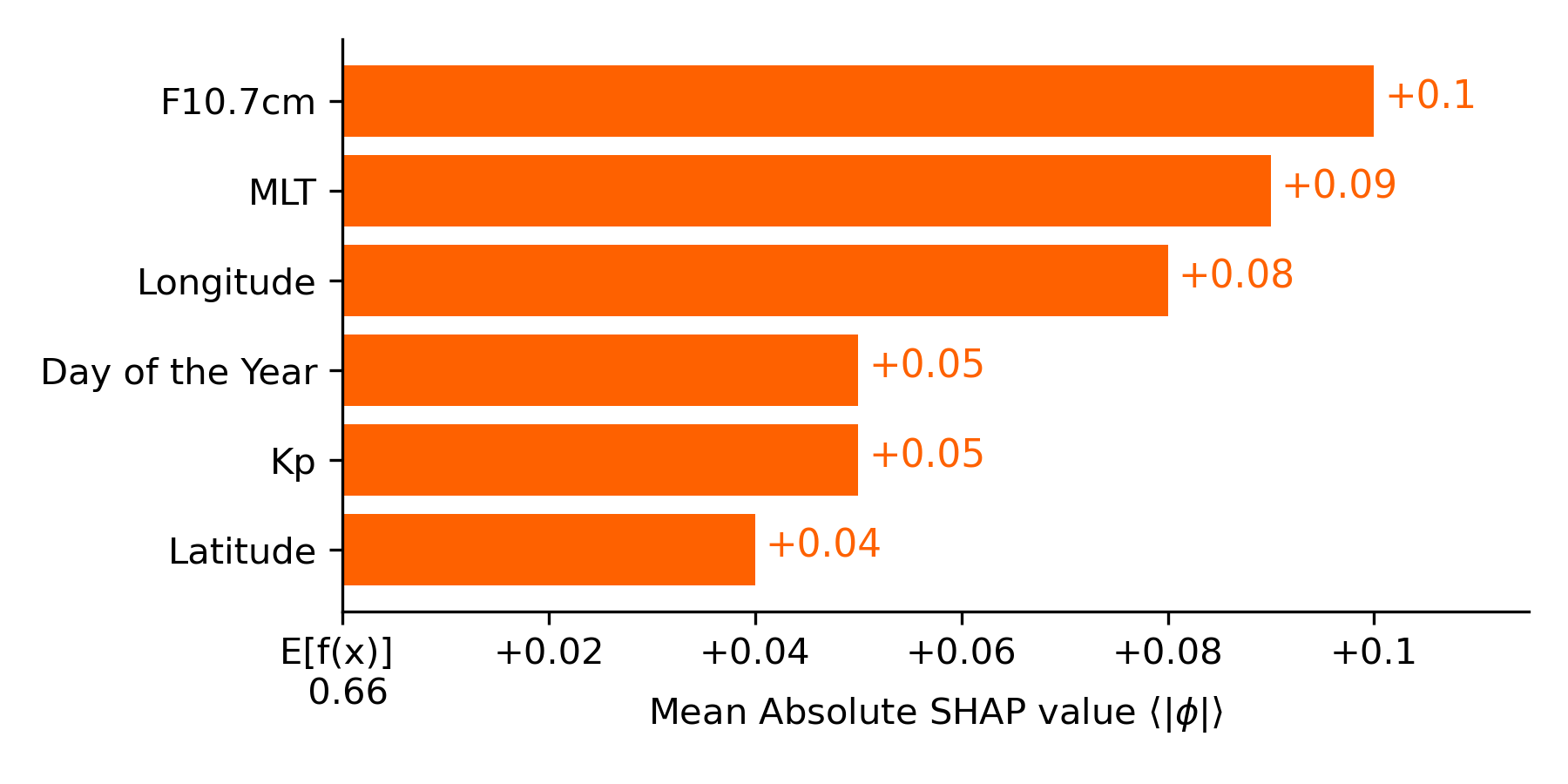}
    \caption{The mean absolute SHAP value across the six features. F10.7 contributes an absolute average of 0.1 to the 0.66 baseline and is the considered the most important feature. Latitude contributes 0.04 to E[f(x)] is considered the least.}
    \label{fig:results_shap-summary}
\end{figure}

Figure \ref{fig:results_shap-summary} shows the mean absolute SHAP value across the six features. It shows that, on average, an F10.7 value will influence the prediction by $0.1$, which is sufficient enough to consider a prediction a `confirmed bubble' \citep{Park2013}. Latitude contributes the least with $\phi = 0.04$. Fig. \ref{fig:results_shap-summary} also shows that F10.7 is the most influential feature, whilst Latitude is the least.

\begin{figure}
    \centering
    \includegraphics[width=\linewidth]{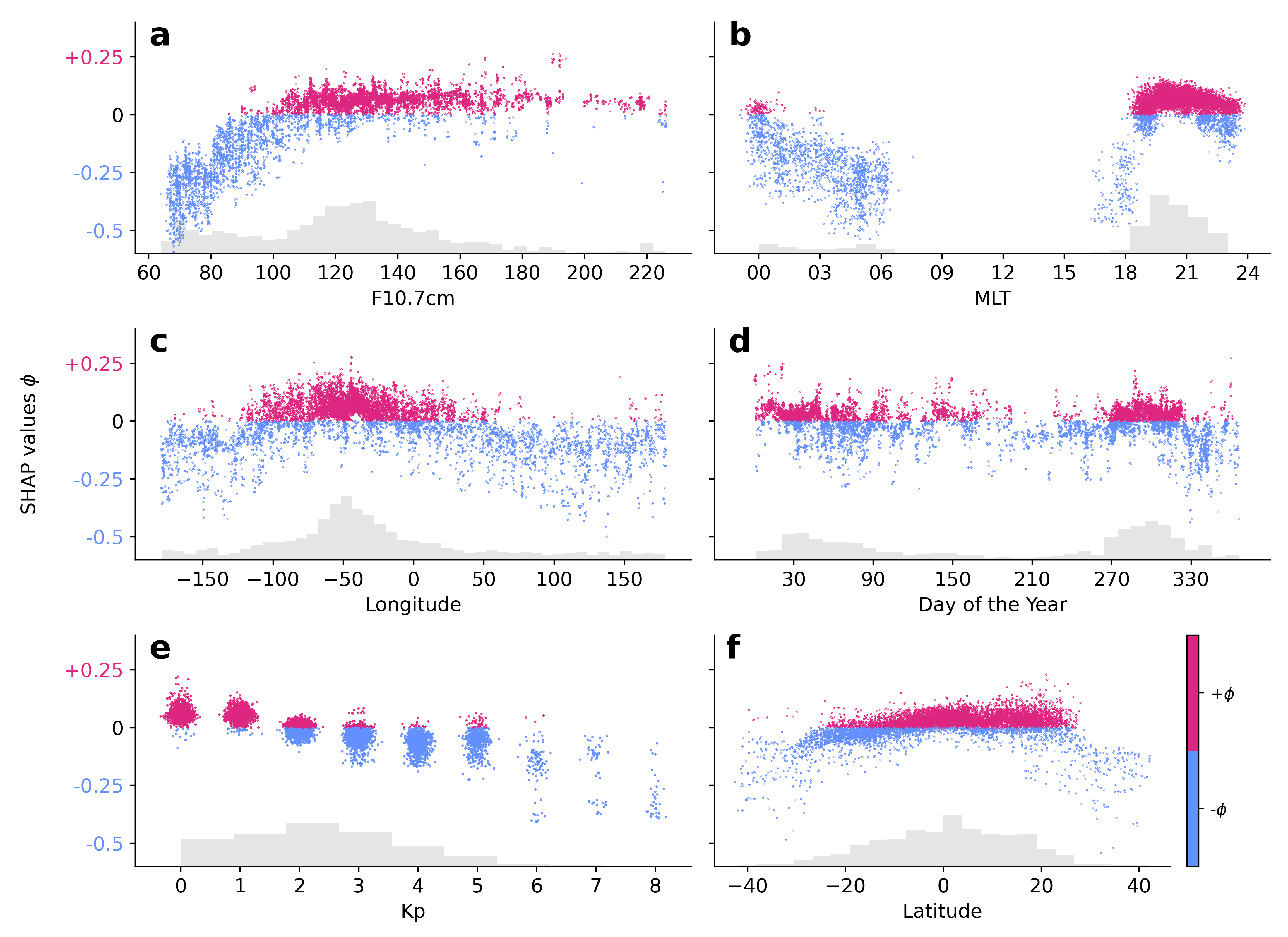}
    \caption{SHAP $\phi$ contributions across the feature space. $\phi > 0$ increases the predicted value of $R^2$, whereas $\phi < 0$ decreases it. Predictions of $R^2 > 0.7$ are considered to be EPBs, so large values of $\phi > 0$ are more likely to be associated with plasma bubbles. Generally the SHAP values follow the climatology outlined in Fig. \ref{fig:ins_clim-hists}.}
    \label{fig:results_shap-scatter}
\end{figure}

We now turn our attention to the feature inputs and corresponding SHAP values. Figure \ref{fig:results_shap-scatter} shows that $\phi$ can be positive and negative, but Eq. (\ref{eq:shap}) means that we can only interpret the contribution to $R^2$ when we take the sum of all the SHAP values. $\phi > 0$ equates to increasing EPB likelihood, whereas $\phi < 0$ is decreasing.

In the F10.7 panel (Fig. \ref{fig:results_shap-scatter}a), we see that low solar activity corresponds to extremely negative SHAP values. This suggests that IBI is primarily detecting magnetic-only fluctuations and that EPBs require F10.7 $>$ 90. 
Secondly, post-sunset values of MLT equate to the highest values of $\phi$, with the contribution peaking at 21 MLT (Fig. \ref{fig:results_shap-scatter}b). It also shows a largely negative contribution after midnight, meaning that most EPBs occur after sunset.
Longitude generally follows the known pattern of increased EPB formation over the American/Atlantic sector (Fig. \ref{fig:results_shap-scatter}c), but there are positive contributions across the longitudinal space.
Unlike the previous features, Day of the Year values generally contribute in a positive and negative way across the entire feature space (Fig. \ref{fig:results_shap-scatter}d). We see high values of $\phi > 0$ around the equinoxes and winter soltice, which is to be expected as EPB formation is generally highest during this period. That said, we also see a high positive cluster around the Earth-Sun perihelion, with the highest value of $\phi$ on Day of the Year = 19. 
Kp provides perhaps the most intriguing insight into EPB climatology (Fig. \ref{fig:results_shap-scatter}e). It clearly shows that increasing Kp equates to negative SHAP values, which reduce the likelihood of an EPB. Beyond Kp $>$ 6 we only see $\phi < 0$ which increases the likelihood of a \textit{B}-only fluctuation. 
Lastly, we see that positive SHAP values are mainly centered around Latitude = 0° to 20° which is expected given EPBs known formation and our use of geodetic coordinates (Fig. \ref{fig:results_shap-scatter}f). 

\begin{figure}
    \centering
    \includegraphics[width=\linewidth]{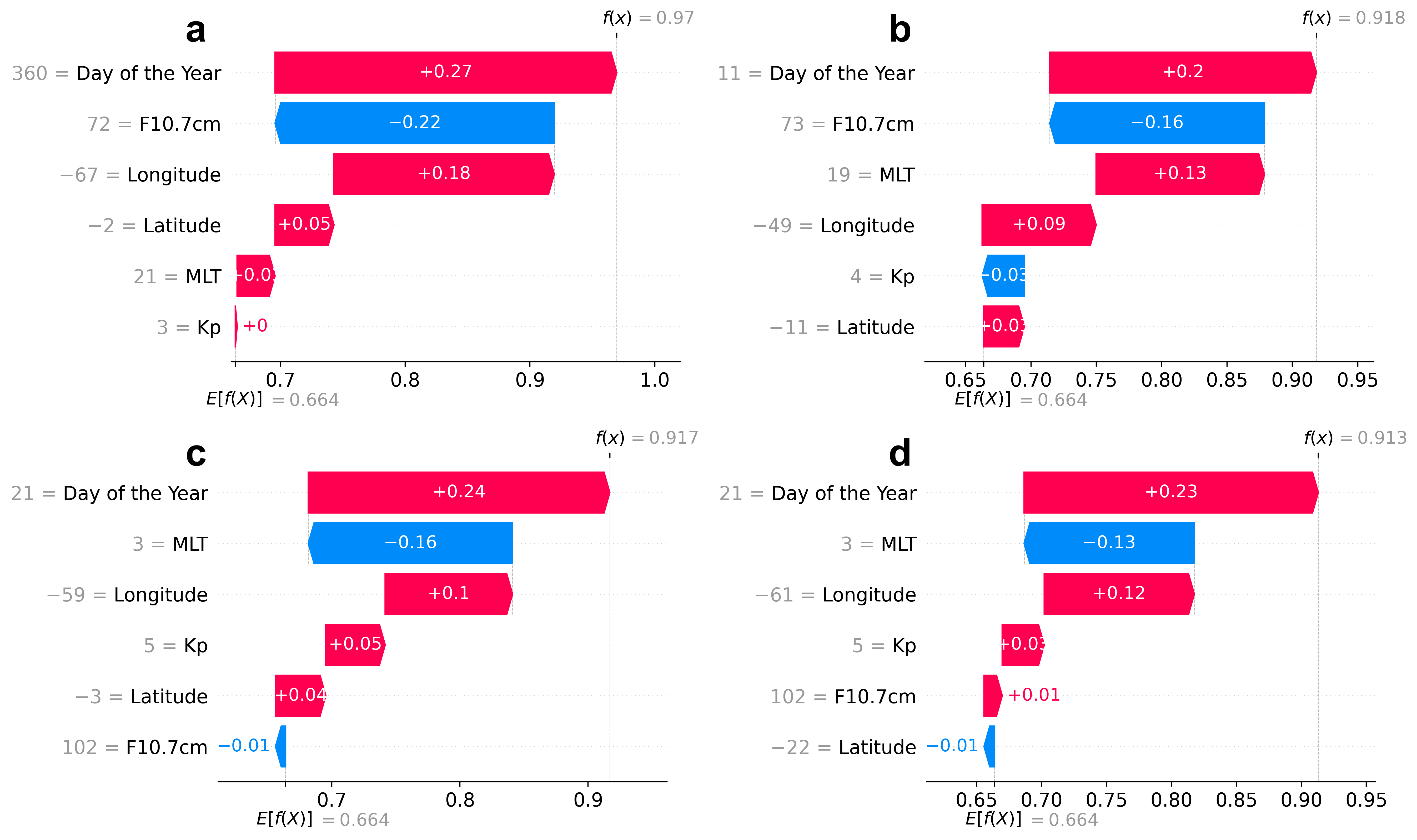}
    \caption{A `waterfall' plot showing 4 predictions around the Earth-Sun perihelion and Kp $>$2. The final prediction value is denoted by $f(x)$, and the values represent the contribution to this from the baseline $E[f(x)] \approx \langle R^2 \rangle = 0.66 $. SHAP ensures that the sum of the contributions always enables a prediction between 0 and 1.}
    \label{fig:results_shap-water}
\end{figure}

Next we examine some of the $\phi > 0$ values at Kp = 4-5 and Day of the Year = 360 to 21. These are intriguing because the former are the \textit{only} positive contributions to EPB prediction during a moderate storm, and the latter exhibits the highest $\phi > 0$ contribution for that feature. Figure \ref{fig:results_shap-water} illustrates the values for Kp and Day of the Year, as well as the other features that contribute to $R^2$. In all cases we see that the IBI value is $>$ 0.9, and is therefore almost certainly an EPB \citep{Park2013}. It's also evident that Day of the Year is the dominant `player', with contributions as high as $\phi = +0.27$ ( Fig. \ref{fig:results_shap-water}a). 
More importantly, Figures \ref{fig:results_shap-water}c-d show the only examples of high Kp equating to positive SHAP values, which also coincide with the Earth-Sun perihelion. Examining this as a whole, Fig. \ref{fig:results_shap-water} shows that a combination of winter solstice / Earth-Sun perihelion, Kp $>$ 2, and low F10.7 equates to a high chance of detecting an EPB.

\section{Discussion}
APE can reliably predict the IBI $R^2$ index on Swarm. If it predicts an $R^2 > 0.7 $ it can be considered an EPB. The model has a high accuracy of \textit{RMSE = 0.08} and exhibits excellent \textit{skill} and \textit{association}. SHAP values reveal the most important features, how features contribute to predictions, and the interrelation between them. We now expand on the IBI observations and SHAP values with respect to geomagnetic activity and seasonal effects. Generally speaking the IBI climate feature observations (Fig. \ref{fig:ins_clim-hists}) and SHAP (Figs. \ref{fig:results_shap-scatter}\&\ref{fig:results_shap-water}) values align with the existing literature: EPBs mainly occur in post-sunset, in the American / Atlantic sector, around the equinox months, and when solar activity is high \citep{Burke2004, Abdu2012, Park2013, Aa2020}. They also show that magnetic-only fluctuations ($R^2 < 0.5$) are more likely post-midnight, during low F10.7, and high Kp.

The above suggests that geomagnetic activity \textit{suppresses} EPB onset. This is supported by the test-set histograms in Fig. \ref{fig:results_shap-scatter}e, which shows that less data \textit{still} results in more positive SHAP contributions when Kp is low. However, geomagnetic activity is both able to \textit{suppress} and \textit{enhance} EPB formation, via DDEF and/or over and under-shielding electric fields \citep{Aa2019, Abdu2012}. Unfortunately, this cannot be fully captured by concurrent Kp owing to DDEF's time-delay effects to the equator. It's possible that indices such as DST or AE are better suited to capturing this, but neither are currently available as forecast products, and thus were excluded from the feature space. To fully capture the influence of geomagnetic activity on EPBs, bespoke indices may be required. For now, this exceeds the remit of this study, especially when the model accuracy is as high as RMSE = 0.08. Kp has been shown to capture day-to-day variability of EPBs during `EPB Season' \citep{Carter2014c}, but additional work is required to capture them during `off-season'. If we assume that F10.7 $<$ 90 to be off-season, then Figure \ref{fig:results_shap-water}c-d, shows Kp could be useful at all times, particularly around the Earth-Sun perihelion. That said, Fig. \ref{fig:results_shap-water}c-d also shows that identical values of F10.7 can have different contributions for different predictions, which shows that interpreting Shapley values requires caution \citep{Kumar2020}. Another interesting feature is that Kp $=$ 5 also coincides with the only $\phi > 0$ at 3 MLT (Figs. \ref{fig:results_shap-scatter}b and \ref{fig:results_shap-water}c-d). EPBs are suppressed after sunset, but enhanced after midnight during large $\Delta$DDEF \citep{Abdu2012}, so these points could be direct evidence of over-shielding effects. That said, the MLT values contribute to the pay-off in a negative way (-$\phi$) and others have reported that over-shielding is more impactful than under-shielding on vertical drifts \citep{Hui2019}, and so more evidence is required to support this.

Turning to the cluster of positive SHAP values around the Earth-Sun perihelion (Fig. \ref{fig:results_shap-scatter}d). We would not expect these $\phi$ values to be higher than the vernal and autumnal equinoxes or December solstice when EPB onset is most probable \citep{Burke2004}. One possible explanation is an increase in the \textit{F} region density around the Gregorian new year, potentially arising from the Earth-Sun perihelion \citep{Rishbeth2006}. The exact cause of this semi-annual variation remains unknown, but we do know that an increased $\sum^F_P$ in Eq. (\ref{eq:sultan-rt}) would increase the growth rate of an EPB \citep{Sultan1996, Carter2020}. This \textit{F} region asymmetry has also been linked to increased atmospheric gravity waves, which are a known seeding mechanism for EPBs \citep{Singh1997, Abdu2009}. That said, the asymmetry happens every year, yet we do not see a large number of points around this period. 
Although further investigation is required into both the seasonal and geomagnetic influences on EPB formation, this discussion highlights the potential of Shapley values to improve our understanding of bubble climatology and predictability.

\section{Conclusions}
In this paper have shown that machine learning can successfully predict the Ionospheric Bubble Index (IBI) on-board the Swarm spacecraft. IBI detects equatorial plasma bubbles in the ionosphere by assessing changes in the plasma density and magnetic field. AI Predictions of EPBs (APE) is able to accurately predict IBI across a range of spatio-temporal conditions. The main findings of our study are summarized below: 
\begin{enumerate}
\item APE fully captures the climatology of EPBs detected by Swarm. This is made possible with the size and resolution of the dataset (9 years @ 1sec), feature selection, and regression-specific model architecture. APE could also be expanded to forecasting as Kp and F10.7 are currently available via NOAA.
\item The XGBoost approach outperforms the other methods (linear regression and random forest) across all metrics. It performs extremely well; presenting a skill, association, and root mean square error score of 0.96, 0.98, 0.08 respectively.
\item APE performs well across the entire feature space, especially post-sunset, in the American/Atlantic sector, around the equinoxes, and when solar activity is high. This is encouraging as most EPBs occur during these periods and locations. Extra consideration may be required when using APE around 3 MLT.
\item SHAP values reveal that F10.7 is the most influential feature, whereas latitude is the least. SHAP values generally align with the existing climatology of IBI EPBs, which validates these results.
\item Additional metrics may be required to fully capture the effects of geomagnetic activity on EPB predictions, but this may compromise APE's ability to forecast them. There is some evidence of high Kp generally suppressing EPB activity, but further investigation into under and over-shielding is required.  
\item The Shapley analysis also reveals that a combination low solar activity, active geomagnetic conditions, and the Earth-Sun perihelion all contribute to increased EPB likelihood. To the best of our knowledge, this is the first time this exact combination of features has been linked to bubble detection. Although its underlying mechanism needs additional investigation, it does showcase the ability of Shapley values to enable new insights into EPB climatology and predictability.
\end{enumerate}

\textbf{This work has now been published \citep{reddy2023predicting}, please use the following reference if you cite the paper:}

Reddy, S. A., C. Forsyth, A. Aruliah, A. Smith, J. Bortnik, E. Aa, D. O. Kataria, and G. Lewis. "Predicting swarm equatorial plasma bubbles via machine learning and Shapley values." Journal of Geophysical Research: Space Physics (2023): e2022JA031183.

\bibliographystyle{unsrtnat}
\bibliography{references}

\end{document}